\documentclass[12pt]{elsart}
\usepackage{graphicx}
\newcommand{\myvec}[1]{\displaystyle {\bf #1}}
\begin{document}
\begin{frontmatter}
\title{Ultra-low energy elastic scattering in a system of three He atoms}
\begin{abstract}
Differential Faddeev equations in total angular momentum representation are used
for the first time to investigate ultra-low energy elastic scattering of a
helium atom on a helium dimer. Six potential models of interatomic
interaction are investigated. The results improve and extend the Faddeev equations based results
known in literature. The employed method can be applied to investigation of
different elastic and inelastic processes in three- and four-atomic weakly
bounded systems below three-body threshold. 
\end{abstract}
\author{V.Roudnev}
\ead{roudnev@cph10.phys.spbu.ru}
\address{Permanent address: 198904 Institute for Physics,
St.Petersburg State University, Uliyanovskaja~1, 
St.Petersburg, Russia\thanksref{author1}}
\thanks[author1]{This work has been performed in the Bogolubov
Laboratory of Theoretical Physics, JINR, Dubna, Russia}
%
%\address{JINR}
\end{frontmatter}
\section{Introduction}
The systems of helium atom small clusters were a subject of a number of
experimental and theoretical researches during the last decades
\cite{Experiment,Experiment2,Experiment3,LM,Az79,Az87,Az91,VanMourik,TTY,Az96,Lim1,Lim2,Pandharipande,Moto1,Moto2,Gloeckle,Carbonell,Fedorov,CPL}.
Development of new experimental techniques
\cite{Experiment,Experiment2,Experiment3} has 
stimulated growing interest to theoretical investigation of bound states and scattering dynamics in
such systems. Availability of numerous potential models 
\cite{LM,Az79,Az87,Az91,VanMourik,TTY,Az96} of He-He interatomic 
interaction provides a good background for such theoretical researches
\cite{Lim1,Lim2,Pandharipande,Moto1,Moto2,Gloeckle,Carbonell,Fedorov,CPL}.

Although the number of papers devoted to investigation of small
helium cluster bound states, especially helium trimer, is considerable, the
number of known results with respect to scattering in such systems is
still very limited. There are only a few estimations for He-He$_2$ scattering
length  \cite{NakaichiMaedaLim,Blume} and,
up to our knowledge, only one group has published results
both  for He-He$_2$ scattering length and phase shifts \cite{Moto1,Moto2}. 
However, the accuracy of the known results for scattering length and phase
shifts seems to be insufficient to resolve the details of different potential
models. 

The calculations presented here are performed on the base of
differential Faddeev equations in configuration space in total angular momentum
representation \cite{TAM}. This formalism was already used in our
calculations of $^4$He$_3$ bound states \cite{CPL,FBS} and allowed us
to obtain very accurate and detailed description for both ground and excited
states of the trimer. Applying this rigorous and reliable formalism to
scattering we expect to obtain benchmark-quality results for 
He-He$_2$ scattering length and phase shifts. The first presentation of these results is
the subject of this Letter.

The paper is organized as follows. In Section~2 we give a brief description of
the equations we solve and a short sketch on the new solution technique. The
results of our calculations are provided with comments and a short discussion 
in Section~3. In Section~4 we give a short resume and present our view on
the possible future application of the methods employed in this paper. 
\section{Formalism}
According to the Faddeev formalism \cite{FaddMerk} the wave function of three particles
is expressed in terms of Faddeev components $\Phi$
\begin{equation}
\label{WFDef}
  \Psi( \myvec {x_1},\myvec{y_1})= 	\Phi_1 ( \myvec {x_1},\myvec{y_1})+
					\Phi_2 ( \myvec {x_2},\myvec{y_2})+
			       		\Phi_3 ( \myvec {x_3},\myvec{y_3}) \: ,
\end{equation}
where $\myvec {x}_{\alpha }$ and $\myvec {y}_{\alpha }$ are Jacobi coordinates 
corresponding to the fixed pair $\alpha$
\begin{equation}
\begin{array}{c}
 \myvec {x}_{\alpha }=(\frac{2m_{\beta }m_{\gamma }}{m_{\beta }+m_{\gamma
 }})^{\frac{1}{2}}(\myvec {r}_{\beta }-\myvec {r}_{\gamma })\ ,\\
 \myvec {y}_{\alpha }=(\frac{2m_{\alpha }(m_{\beta }+m_{\gamma })}{m_{\alpha
 }+m_{\beta }+m_{\gamma }})^{\frac{1}{2}}(\myvec {r}_{\alpha }-\frac{m_{\beta
 }\myvec {r}_{\beta }+m_{\gamma }\myvec {r}_{\gamma }}{m_{\beta }+m_{\gamma }})
 \ .
\end{array}
\label{Jacoord}
\end{equation}
Here $\myvec {r}_{\alpha }$ are the positions of the particles in the center-of-mass
frame. The Faddeev components obey the set of three equations
\begin{equation}
\label{eqf}
\begin{array}{rcl}
 (-\Delta _{x}-\Delta _{y} +V_{\alpha }(\myvec{x}_{\alpha }) & - & E)\Phi _{\alpha }(\myvec {x}_{\alpha },\myvec {y}_{\alpha })  =  \\
 & = & \displaystyle
 -V_{\alpha }(x_{\alpha })\sum _{\beta \neq \alpha }\Phi _{\beta }(\myvec {x}_{\beta },\myvec {y}_{\beta })
  \\ 
  \alpha=1,2,3  
\end{array} \, ,
\end{equation}
where $V_{\alpha }(\myvec{x}_{\alpha })$ stands for pairwise potential.
To make this system of equations suitable for numerical calculations
one should take into account the symmetries of the physical system. 
Exploiting the identity of Helium atoms one can reduce the system of equations (\ref{eqf}) 
to one equation \cite{FaddMerk}. 
Since all the model potentials are central it is also possible to factor out the
degrees of freedom corresponding to 
the rotations of the whole cluster \cite{TAM}. For the case of zero total angular momentum 
the reduced Faddeev equation reads
\begin{equation}
\label{Fadd3}
 \displaystyle
  (H_0+ V(1+P)-E)\Phi (x,y,z)=0\: ,
\end{equation}
where $H_0$ is the restriction of free Hamiltonian to the intrinsic space
corresponding to zero total angular momentum
\[
H_0=-\frac{\partial ^{2}}{\partial x^{2}} 
   -\frac{\partial ^{2}}{\partial y^{2}}
   -(\frac{1}{x^{2}}+\frac{1}{y^{2}})
   \frac{\partial }{\partial z}(1-z^{2})
   \frac{\partial }{\partial z} \ , 
\]
$x$, $y$ and $z$ are the intrinsic coordinates corresponding to the one
selected partitioning of the three particles into 2+1 subsystems
\begin{equation}
\label{intrcoord}
  x=|\myvec {x}| \ , \, y=|\myvec {y}|\ , \, \displaystyle 
  z=\frac{(\myvec {x},\myvec {y})}{xy} \, ,
\end{equation}
$V=V(x)$ is the two-body potential and $P$ is an operator defined as
\[
  P \Phi(x,y,z)=xy(\frac{\Phi(x^+,y^+,z^+)}{x^+y^+} +
  \frac{\Phi(x^-,y^-,z^-)}{x^- y^-}) \ .
\]
Here $x^{\pm}$, $y^{\pm}$ and $z^{\pm}$ are Jacobi coordinates
corresponding to other partitionings of three particles into
subsystems. Explicit expressions for these coordinates in the case of
particles with equal masses are given by the following formulae
\[
\begin{array}{l} \displaystyle
x^{\pm }=(\frac{x^{2}}{4}+\frac{3y^{2}}{4}\mp
\frac{\sqrt{3}}{2}xyz)^{1/2} \; ,\\  \displaystyle
y^{\pm }=(\frac{3x^{2}}{4}+\frac{y^{2}}{4}\pm
\frac{\sqrt{3}}{2}xyz)^{1/2}\; ,\\  \displaystyle
z^{\pm }=\frac{\pm \frac{\sqrt{3}x^{2}}{4}\mp
\frac{\sqrt{3}y^{2}}{4}-\frac{1}{2}xyz}{x^{\pm } y^{\pm }}\; .
\end{array}
\]

In this paper we concentrate our attention on the case of elastic scattering
only. In this case the solution of the Faddeev equation (\ref{Fadd3}) can be
presented as a sum of two terms 
\begin{equation}
  \Phi (x,y,z)=\chi(x,y,z)+\tilde{\Phi}(x,y,z) \ .
\label{abc0}
\end{equation}
The function $\chi(x,y,z)$ corresponds to the initial state of the system, i.e.
free motion of the atom and the dimer, the second term $\tilde{\Phi}(x,y,z)$
corresponds to the scattered state of the atom and the dimer. Explicit expression for
$\chi(x,y,z)$  reads
\[
  \chi(x,y,z)=\varphi _{2}(x)\frac{\sin ky}{k} \ ,
\]
where $\varphi _{2}(x)$ is the two-body bound state wave function, $k=\sqrt{E-E_{2}}$,
$E_{2}$ is the energy of the two-body bound 
state and $E$ is the total energy of the three-body system in the center of mass frame.
The  asymptotic boundary
condition for the function $\tilde{\Phi}(x,y,z)$ is defined as
\begin{equation}
 \tilde{\Phi}(x,y,z) \mathop{\longrightarrow }\limits_{y \rightarrow \infty} \: \frac{a(k)}{k} \varphi _{2}(x)\cos ky \: ,
 \label{abc1}
\end{equation}
where $a(k)$ stands for the elastic scattering amplitude. Substituting the
representation (\ref{abc0}) to the equation (\ref{Fadd3}) we get the equation for the
function  $\tilde{\Phi}(x,y,z)$
\begin{equation}
\label{Fadd4}
\displaystyle
  (H_0+V(1+P)-E)\tilde{\Phi} (x,y,z)
   =-VP \chi(x,y,z) \ .
\end{equation}

To calculate the low-energy scattering characteristics of the system
one has to solve the equation (\ref{Fadd4}) with the asymptotic boundary condition
(\ref{abc1}). The scheme of numerical solution of the equation is not the
subject of the Letter, but we would like to mention some features of the method we
employ that seem to be original or at least that were not applied for few-body calculations 
previously. 

The general scheme of the problem discretization reproduces the
one used in our helium trimer calculations \cite{CPL,FBS}. However, the way we use
tensor-trick preconditioning  \cite{Groning} is changed, that allowed us to
reduce the dimension of the corresponding linear problem almost twice. Due to
the structure of the Faddeev equations, in the case of 
potentials decreasing sufficiently fast, the region in
configuration space where an interaction between atoms vanishes has a simple
geometric shape of a hypercylinder $x>x_{max}$ exterior. Therefore, outside of the
hypercyllinder the Faddeev component $\tilde{\Phi} (x,y,z)$ satisfy an equation for
free particles. At the same time, for large values of $y$, where the component meets
the asymptotic boundary conditions, it satisfies the equation
\begin{equation}
\label{asymp}
  (H_0 +V(x)-E)\tilde{\Phi} (x,y,z) \approx 0 \ , y>y_{max} \ .
\end{equation}
Taking into account the aforementioned observations one can find that a 
new component defined as
\begin{equation}
\tau (x,y,z) \equiv (H_0 +V(x)-E)\tilde{\Phi} (x,y,z) 
\label{phidef}
\end{equation}
is localized much better in coordinate $x$  than the original component
$\tilde{\Phi} (x,y,z)$ and can be approximated using much less grid points. 
One can even prove that unlike the original component $\tilde{\Phi} (x,y,z)$ 
such component is square integrable for scattering states (square
integrability of a similar object was briefly 
discussed in \cite{PappYakovlev}). The equation for $\tau(x,y,z)$ reads
\begin{equation}
  \tau (x,y,z)
   =-V P \chi(x,y,z)-V P(H_0+V-E-i0)^{-1}\tau (x,y,z) \: .
\label{tauEq}
\end{equation}
This is the equation which we solve numerically using Krylov subspace projection
techniques.
The operator $(H_0+V-E-i0)^{-1}$ entering the equation is a resolvent of so
called cluster Hamiltonian which corresponds to a system of one free and two
interacting particles.  In our approach this operator is constructed by solving
the corresponding differential equation with appropriate asymptotic boundary
conditions, what is made using tensor-trick technique. 
Having the equation  (\ref{tauEq}) solved, we recover the
original component $\tilde \Phi (x,y,z)$ according to the definition 
(\ref{phidef}), using the same tensor-trick.

As a result of numerical procedure we get an approximate solution
$\tilde \Phi (x,y,z)$. To recover the scattering amplitude we compare the
numerical solution with the asymptotic representation (\ref{abc1}) pointwise.
This way we define the function 
\[
  a(k;x,y,z)\equiv \frac{k\tilde{\Phi}(x,y,z)}{\phi _{2}(x)\cos ky} \ .
\] 
Stability of this function with respect to variations  of $x$, $y$ and $z$ coordinates
indicates, that the asymptotic region is reached in the calculations and the function
approaches the scattering amplitude. Such simple test for an 
asymptotic region is made possible by use of Cartesian coordinates \cite{Carbonell}.
Another advantage provided by the usage of Cartesian coordinates is a simple test
for the quality of a grid taken in $x$ coordinate: this grid should support the
correct value of the dimer binding energy \cite{Carbonell,CPL}. 

\section{Results}
In this section we describe our calculations of He-He$_2$ elastic
scattering and give a brief comparison of our calculations with other published
results. 

On the preliminary stage of the calculations we have verified, that our code
reproduces the known properties of helium trimer ground and excited states for a
particular potential model. To do it, a special version of the code using exactly
the same procedures as the ones involved in scattering calculations was
developed. All the known figures of  trimer ground and excited state energies
\cite{CPL} were successfully reproduced. 
After that we have performed a set of calculations with maximal values
of $x$ and $y$ taken at 2500~\AA. The analysis of the results has allowed us to
conclude that for small values of atom-dimer kinetic energy the asymptotic region
starts approximately from 1000~\AA \ in $y$ coordinate. For all the values of $y$ 
taken in the region $y>1000$~\AA \  and for 
all $x>2$~\AA \ (outside of the repulsive core region) the value of scattering
amplitude was stable up to 5 figures. The most of the subsequent calculations were
performed with maximal  values of $x$ and $y$ fixed at 1200~\AA. To ensure that
our results are stable and accurate we have performed the 
calculations of the He-He$_2$ scattering length, phase shifts and scattering
amplitude using different sets of grids. The stability of the scattering length
and scattering amplitude with respect to the number of grid points is
demonstrated in the Tables \ref{tabLstab} and \ref{tabAstab}. The presented data allows to
estimate the accuracy of our scattering length calculations as low as
$0.1$\% what is comparable with the accuracy of the used physical constants.

In the Table \ref{mainRes} we present the results of scattering length and phase
shifts calculations for six different potential models. The potentials HFD-B,
LM2M2 and TTY were already used in atom-dimer scattering calculations
\cite{Moto1,Moto2,Blume}. The estimations of scattering observables for more recent SAPT, SAPT1
and SAPT2 potentials are presented here for the first time. The potentials SAPT1 and SAPT2
\cite{Az96} are  
constructed taking into 
account retardation effects that lead to the change of the highest term from
$O(x^{-6})$ to $O(x^{-7})$ at large interatomic distances \cite{Dalgarno}. All
other potentials have $O(x^{-6})$ long-distance asymptotic behavior. 

To prove the correctness of our calculations we must compare our results with
known published results and to explain the differences between our results and
results of other authors if possible. Unfortunately, only small number of
independent results were published with respect to He-He$_2$ scattering 
problem. The number of calculations that take into account the interatomic
interactions in the states with non-spherical symmetry is limited, up to our
knowledge, only by the results of Motovilov et al. \cite{Moto1,Moto2} and
recent calculations of Blume et al. \cite{Blume}. Our results for scattering
length are brought together with the results known from literature in the 
Table~\ref{compar}. Obviously, the difference between our results and other results is
far from negligible and requires an explanation. 

\begin{figure}[p]
{ %\vspace{7mm}
\centering \includegraphics[width=0.7 \textwidth]{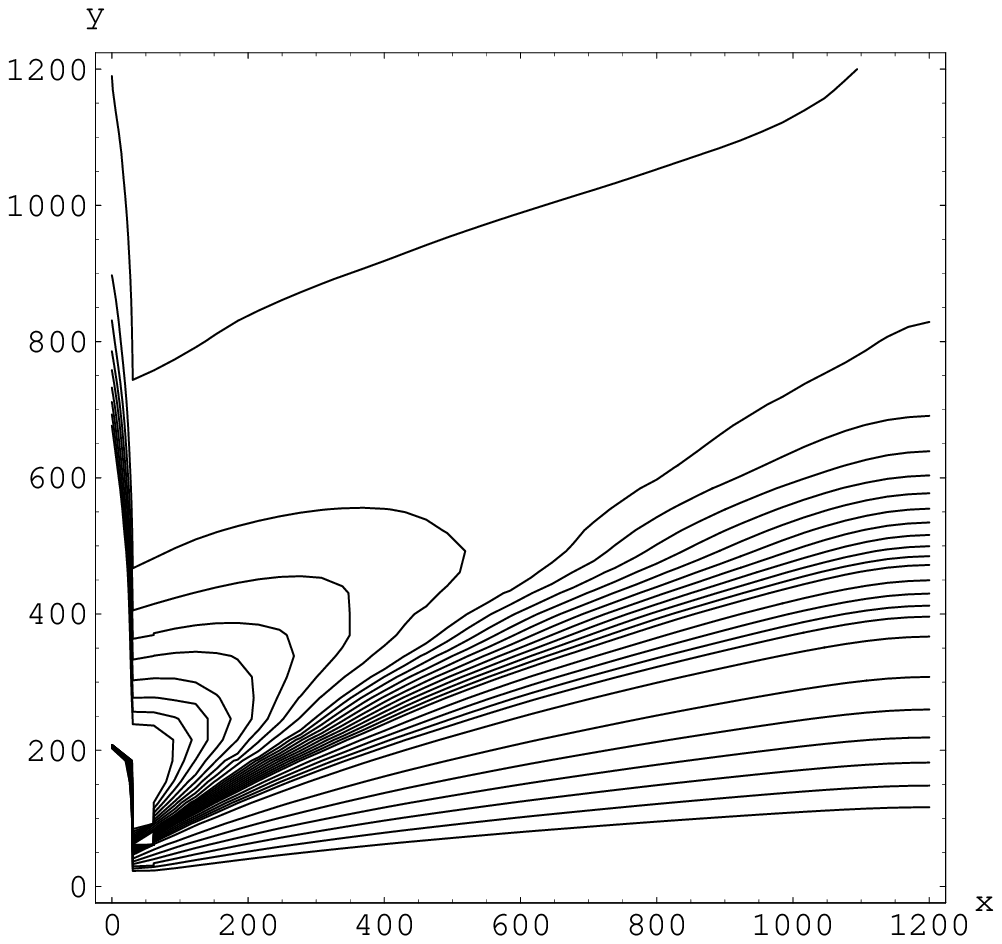} \par}
\caption{Contour plot of $\displaystyle \lim_{k\rightarrow 0}a(k;x,y,0)/k$,
          the values of $x$ and $y$ are given in \AA \label{ampCont}}
\end{figure}

The results of Motovilov et al. \cite{Moto1,Moto2} were obtained by
solving Faddeev differential equations in bispherical harmonics 
representation, and only the first three
terms of the bispherical expansion were taken into account. It can be shown, that this approach
corresponds to the simplest grid in $z$ coordinate that can be used within our
method. This grid consists of the only one interval and the corresponding spline
basis contains 6 basic polynomial functions. However, as one can see from the 
Table~\ref{tabLstab}, restriction of our basis to the simplest case can not cover
the observed difference of 10\% in scattering length. The source of the observed
discrepancy can be found in the cutoff distances used in the numerical
calculations. The values reported in \cite{Moto2} were obtained for the
cutoff radius of 460~\AA. According to \cite{MotovilovPC} the choice of this cutoff
radius was forced by the limitations of the available computer facilities. To check
the version about strong influence of the cutoff radius to the result we have
performed calculations with cutoff parameters $x_{max}=800$~\AA \ and
$y_{max}=460$~\AA. The results of these calculations are also presented in the
Table~\ref{compar}. Evidently, they are in perfect agreement with the results of
\cite{Moto2}. Therefore we can confidently confirm the consistency of our
results and the results published by Motovilov et al.~\cite{Moto2}. 
However, we must note, that for such reduced cutoff parameters the scattering amplitude
varies within 30\% interval even outside of the core region what indicates that such
small cutoff distance is not enough to obtain stable results. As one can see from the
Fig.~\ref{ampCont}, the asymptotic region where the scattering amplitude is rather stable with
respect to variations of coordinates starts approximately from
$y\approx 1000$~\AA.

\begin{figure}[p]
{ %\vspace{7mm}
\centering \includegraphics[width=0.7 \textwidth, angle=270]{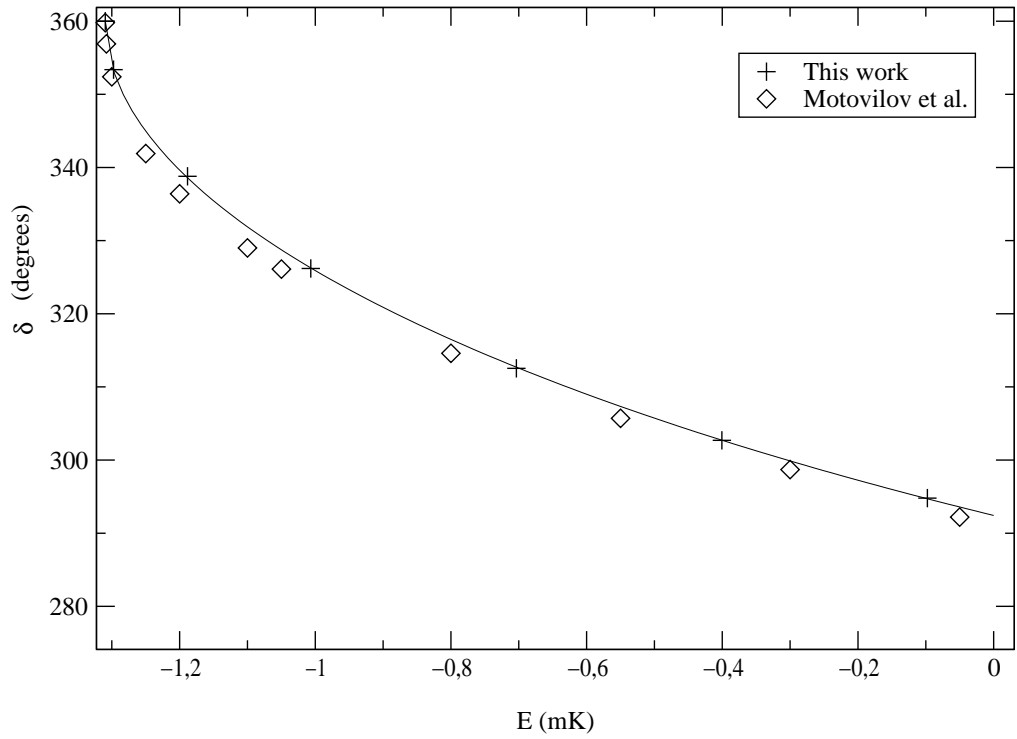} \par}
\caption{He-He$_2$ scattering phase shifts for TTY potential. The energy $E$ is
the energy of the system in the center of mass frame. \label{FigPhase}}
\end{figure}

Even though the results for scattering length published in \cite{Moto2} considerably differs from
the results presented in this Letter,
the discrepancy between the results for phase shifts, as one can see from the
Fig.~\ref{FigPhase}, is not so big. The difference slightly grows towards 2-body
threshold, however in the vicinity of 3-body threshold ($E=0$) the difference is
rather small.

Unfortunately, the estimations of numerical error for the calculations performed by
Blume et al. \cite{Blume} are not available. Being aimed to calculation of larger clusters,  their
technique differs much from the one employed in this work. 
It makes detailed comparison of our results difficult, and we can only
mention that their result for He-He$_2$ scattering length is 
in better agreement with the our one than the result of \cite{Moto2}.

%%%%%%%%%%%%%%%%%%%%%%%%%%%%%%%%%%%%%%%%%%%%%%%%%%%%%%%%%%%%%%%%%%%%%%%%%%%%%%%%%%%%%%%%%%%%%%%%%%%%
\begin{table}[p]
\caption{Convergence of the He-He$_{2}$ scattering length with respect
to the number of grid points (HFD-B potential)
\label{tabLstab}}
\vspace{5mm}
\begin{center}
\begin{tabular}{|l|c|c|c|}
\hline 
Grid               	& $\times$ 6$_z$   	&  $\times$ 9$_z$  	& $\times$ 15$_z$  \\
\hline 
\hline 56$_x$ $\times $ 86$_y$  	&   122.31 	&  121.91	&  121.93 \\
\hline 68$_x$ $\times $ 101$_y$   	&   122.32 	&  121.91	&  121.92 \\
\hline 77$_x$ $\times $ 116$_y$  	&   122.34 	&  121.94	&    -	  \\
\hline 86$_x$ $\times $ 131$_y$   	&   122.32 	&  121.90	&    -	  \\
\hline 98$_x$ $\times $ 146$_y$  	&   122.31 	&  	-	&    - 	  \\
\hline 
\end{tabular}
\end{center}
\end{table} 
%%%%%%%%%%%%%%%%%%%%%%%%%%%%%%%%%%%%%%%%%%%%%%%%%%%%%%%%%%%%%%%%%%%%%%%%%%%%%%%%%%%%%%%%%%%%%%%%%%%%
\begin{table}[p]
\caption{Convergence of the He-He$_{2}$ scattering amplitude with respect
to the number of grid points (HFD-B potential) at $E_{rel}=E-E_2=1.515$~mK 
\label{tabAstab} }
\vspace{5mm}
\begin{center}
\begin{tabular}{|l|c|c|c|}
\hline 
Grid               	& $\times$ 6$_z$  	&  $\times$ 9$_z$  	& $\times$ 15$_z$  \\
\hline 
\hline 56$_x$ $\times $ 86$_y$  	&  -3.023 	&  -2.996	& -2.997   \\
\hline 68$_x$ $\times $ 101$_y$   	&  -3.036  	&  -3.009	& -3.009  \\
\hline 77$_x$ $\times $ 116$_y$  	&  -3.041 	&  -3.014	&    -	\\
\hline 86$_x$ $\times $ 131$_y$   	&  -3.041  	&  -3.013	&    -	\\
\hline 98$_x$ $\times $ 146$_y$  	&  -3.039  	&  	-	&    -	\\
\hline 
\end{tabular}
\end{center}
\end{table} 
%%%%%%%%%%%%%%%%%%%%%%%%%%%%%%%%%%%%%%%%%%%%%%%%%%%%%%%%%%%%%%%%%%%%%%%%%%%%%%%%%%%%%%%%%%%%%%%%%%%%%
\begin{table}[p]
\caption{Scattering lengths and phase shifts for different potential models \label{mainRes}}
\vspace{5mm}
\begin{center}
\begin{tabular}{|l|c|c|c|c|c|c|c|c|}
\hline 
Potential & $l_{sc}$ (\AA) &  \multicolumn{7}{c|}{$\delta$ (degrees) for different values of
$E_{rel}=E-E_2$ (mK) } \\
          &                & \parbox{0.1 \textwidth}{ 0.01212} 
	                              &  \parbox{0.1 \textwidth}{ 0.1212}  
				                   &  \parbox{0.1 \textwidth}{ 0.303}
						                & \parbox{0.1 \textwidth}{ 0.606} 
								           & \parbox{0.1 \textwidth}{ 0.909}
									              & \parbox{0.1 \textwidth}{ 1.212} 
										                 &  \parbox{0.1\textwidth}{1.515} \\
\hline 
 HFD-B 	  &  121.9         &  353.01  &   337.95   &   325.37   &  311.86  &  302.2   &  294.6   &  288.3    \\
 LM2M2    &  115.4         &  353.38  &	  338.79   &   326.19   &  312.54  &  302.7   &  294.8   &  - \\%288.3    \\
 TTYPT    &  115.8         &  353.35  &	  338.73   &   326.13   &  312.47  &  302.6   &  294.3   &  - \\%286.5    \\
 SAPT     &  123.7         &  352.91  &	  337.72   &   325.10   &  311.66  &  302.0   &  294.4   &  288.2   \\
 SAPT1    &  122.4         &  352.99  &	  337.90   &   325.30   &  311.84  &  302.2   &  294.6   &  288.3   \\
 SAPT2    &  123.1         &  352.95  &	  337.81   &   325.20   &  311.75  &  302.1   &  294.5   &  288.3   \\
\hline 
\end{tabular}
\end{center}
\end{table} 
%%%%%%%%%%%%%%%%%%%%%%%%%%%%%%%%%%%%%%%%%%%%%%%%%%%%%%%%%%%%%%%%%%%%%%%%%%%%%%%%%%%%%%%%%%%%%%%%%%%%
\begin{table}[p]
\caption{Comparison of calculated scattering length with the results known from literature \label{compar}}
\vspace{5mm}
\begin{center}
\begin{tabular}{|l|c|c|c|c|}
\hline 
Potential & \cite{Moto2} &  \cite{Blume} & \multicolumn{2}{|c|}{This work}   \\
          &                      & 		 & \parbox{0.2 \textwidth}{ $x_{max}=y_{max}=1200$\AA} &    
	                                           \parbox{0.2 \textwidth}{$x_{max}=800$\AA, $y_{max}=460$\AA} \\
\hline HFD-B 	&   135 $\pm$ 5 	 &  N/A	         & 121.9 $\pm$ 0.1 & 132       \\
\hline LM2M2  	&   131 $\pm$ 5 	 &  126	         & 115.4 $\pm$ 0.1 & 134       \\
\hline 
\end{tabular}
\end{center}
\end{table} 
%%%%%%%%%%%%%%%%%%%%%%%%%%%%%%%%%%%%%%%%%%%%%%%%%%%%%%%%%%%%%%%%%%%%%%%%%%
%%%%%%%%%%%%%%%%%%%%%%%%%%%%%%%%%%%%%%%%%%%%%%%%%%%%%%%%%%%%%%%%%%%%%%%%%%
%%%%%%%%%%%%%%%%%%%%%%%%%%%%%%%%%%%%%%%%%%%%%%%%%%%%%%%%%%%%%%%%%%%%%%%%%%
\section{Conclusions}
New results for He-He$_2$ scattering length and elastic
scattering phase shifts are presented for six different potential models. Being based
on a rigorous theoretical approach suitable for atom-dimer scattering calculations
(Faddeev equations in total angular momentum representation, 
Cartesian coordinates) and an original numerical technique, these
results improve the reference results of Motovilov et al.\cite{Moto2} by two
orders of magnitude. The results for SAPT, SAPT1 and SAPT2 potential models
are new.  

Comparison of elastic scattering parameters calculated for the available potential models shows, that
difference between the parameters is rather small and can hardly reveal the fine details of the
He-He interatomic interaction. It gives promise that low-energy properties of He-He interaction can
be reproduced within much simpler model than realistic potential models
employed in this work.

The numerical technique developed and applied in this work can be further applied in investigation
of bound states and scattering of other exotic systems. For instance, it can be applied to calculate
bound states and scattering in the systems of other rare gas atoms. Four-body calculations based on
Faddeev-Yakubovsky differential equations \cite{FRYV} can also be performed within our approach.

\section*{Acknowledgements}
The author is like to express his special gratitude to Dr.~A.K.Motovilov for
attracting his attention to the subject and for a recommendation to visit the Bogolubov
Laboratory of Theoretical Physics, JINR, Dubna, where the final part of this
work was performed in the framework of JINR-UNESCO agreement.
The author is grateful to Prof.~S.L.Yakovlev for constant moral support and useful practical advises, 
%to Prof.~S.I.Vinitsky for reading preliminary version of the paper and criticism, 
to Dr.~D.E.Monakhov for organising the visit to Dubna,
to Dr.~F.M.Pen'kov for encouraging discussions 
and to I.M.Alexandrovitch for inspiration and warm hospitality during the stay in Dubna.

\end{document}